\documentstyle{article}
\newcommand{\deriv}[2]
{\!\!\stackrel{\scriptstyle [#2]}{\strut#1}\!\!{}}
\newcommand{\tdots}[1]
{\:\lefteqn{\ddot{\lefteqn{\vphantom{#1}}}}\hspace{2.8pt}%
\dot{\lefteqn{\vphantom{#1}}}\!\!\!#1}

\begin{document}
\large

\author {G.S.Bisnovatyi-Kogan$^1$, \and S.V.Repin$^1$}
\title{\bf Timing of gamma-ray pulsars: search in seven - parametric space}
\date{}
\maketitle

\centerline { $^1$ Space Research Institute, 84/32, Profsoyuznaya st.,
             Moscow, 117810, Russia.}

\begin{abstract}
Timing of Geminga gamma-ray pulsar is done using data of COS B and EGRET.
It is shown, that errors in angular coordinates of sources similar to
Geminga strongly influence a determination of $\ddot \nu$, so that
at angular precision less then $10^{-3}$ arc sec determination of
the value of $\ddot \nu$ by means of criterion gives error more then $100 \%$.
Attempt have been done to improve coordinates of a gamma-ray pulsar using
timing analysis. In
addition to search of $\nu$, $\dot\nu$ and $\ddot\nu$, a technique is first
developed permitting search of two angular coordinates, absolute speed
value and direction of a proper motion. In that way timing of gamma-ray
pulsars gives amount of information compatible with radiopulsars, but
using of real data gives much poorer precision. In gamma-ray sources with
rare pulses the periodicity criteria are quite different from the ones in
radio region. Data on the coordinates and proper
motion of Geminga, obtained from timing studies, do not contradict
inside the errors to its identification with $G^{\prime \prime}$
star and its
proper motion. These errors are larger then ones in optical measurements,
but are smaller then corresponding errors in $X$-ray and $\gamma$-ray
data. Estimations of the gamma-ray pulsar coordinates and its proper motion
could be obtained independently on its optical or radio component, and are
available in their absence.
\end{abstract}

\section {Introduction}

    The discovery of Geminga as a ``true'' pulsar, but without visible
radioemission, gave additional evidence to the idea that hard
gamma-ray emission ($E\ge$ 30 MeV) is an inherent property of
pulsar radiation. Before this only the Vela pulsar, the strongest hard
gamma-ray source and the young Crab pulsar gave hints of this possibility.
Observations on the sky, made by EGRET on CGRO
\cite{thompson1} \cite{thompson2}
have shown that only young pulsars with age not exceeding several
tens thousand years and, possibly, millisecond pulsars \cite{ver}
give observable flux in gamma radiation. It is
not yet clear, what is the mechanism of this gamma radiation and
whether it has a
threshold character, or if there is a gradual decrease of hard
gamma ray flux with an age.

   The existence of the Geminga pulsar indicates, that there could be other
gamma ray pulsars with no radioemission, which are exhibited in EGRET
observations as ordinary point-like sources, see i.g.
\cite{Mukher,braz2}.

     Determination of pulsations in a hard gamma ray source is a very
difficult problem, connected with rareness of arriving quanta
$\delta t \gg P$, and small total number of quanta. When the value
of the period is known from other observations (radio or $X$-ray),
timing analysis gives the possibility to reproduce this periodicity also
in gamma region \cite{bignami1}, \cite{egretdat}.
When there is no information about the period, it could, in principle, be
found from pure gamma data \cite{Gur1}, but this could
take enormous amount of computer time and has not been in full realized in
practice.

     A position on the sky of pulsars with no radioemission cannot be
established precisely; the best position obtained from $X$-ray
observations are between $3^{\prime\prime}$ (Einstein) and
$5^{\prime\prime}$ (ROSAT) for 90\% level \cite{becker,becker1}.
Optical identification of
Geminga with very faint $> 25^m\!\!\!.\,5$ object have been done in
\cite{bignami3} and later measurements of its proper motion
\cite{bignami4} and parallax \cite{carav} can be considered
as an evidence of reality of this
identification.

     Timing of Geminga in hard gamma region based on COS--B
\cite{bignami1}, \cite{circ5541} and EGRET \cite{egretdat} data
gave anomalously high braking index
$n = \nu\ddot\nu / \dot\nu^2 \sim 10-30$,
corresponding to very high second derivative $\ddot\nu$.  While there is
a possibility, that it is connected with poor precision of $\ddot\nu$
determination, it is worth to investigate other explanations. It was
suggested in \cite{bispropmov}, that high value of $n$ ($\ddot\nu$)
results from errors in its coordinates, leading to incorrect barycenter
reduction procedure, which spoils the timing procedure. This problem is
well known for pulsar timing, where the error in coordinates give a
one year periodical deviations from the smooth curve in the pulse
arriving time, what permits to improve pulsar position to amazing
precision of the order and even better than VLBI observations
\cite{harrison,taylor}.

     Here we describe a method for investigation of timing of
gamma pulsars, represented by periodical objects with
rare pulses, which gives possibility to determine 7 parameters of
a gamma pulsar: frequency $\nu$, its two derivatives $\dot\nu$ and
$\ddot\nu$, angular coordinates $\alpha$ and $\delta$ of the source,
absolute value $v$ and direction of a velocity of a proper motion,
characterized by an angle $\theta$. When registered pulses are rare, so
that their time separation $\delta t$ is much larger then the period
$P$, the method of investigation is  quite
different from the one, used for radiopulsars. On the artificial
sample of data, which properties simulate Geminga, but in contrary,
have very narrow light curve ($\delta$-function),
no systematic errors and no false quanta, we have managed to determine
coordinates and proper motion parameters with very high precision. This
precision is decreasing when we go to pulses with a finite width, in
presence of background and systematic errors.

     Application of this method to real data sample of Geminga from
COS--B and EGRET was not so successful, because of
smooth light curve, ``nonperfectness''
of data, possible glitches.
We present here results for most probable position and proper
motion characteristics, determined by using of periodicity criteria,
which are not in contradiction with more precise optical data, and has
better coordinate precision then $\gamma$ - ray or $X$ - ray data.
Correlation properties of timing criteria, used for sources with
rare pulses are investigated, before applying them to a timing
procedure.

\section
{Barycentric corrections: account for angular coordinates correction
and proper motion}

For timing analysis all data must be presented in the same coordinate
system, which as a rule is connected with a barycenter of the Solar
system. Consider first a situation, when angular coordinates of a
source $\alpha$ and $\delta$ are known exactly.
On Fig.~4
$xyz$ is a coordinates system, that remains
at rest with respect to distant stars. The point O is the barycenter
of the Solar system.
The space probe is in the point S with Cartesian coordinates
$(x_0,y_0,z_0)$.
A direction to a source is defined by a straight line with coordinate
angles $\alpha$ and $\delta$, where
$\alpha$ angle is counted in $xy$ plane counterclockwise from the
positive direction of $x$-axis. An angle $\delta$ is countered in a plain,
perpendicular to $xy$ plane.
$SB$ is a perpendicular from the space
probe (point $S$) onto a line from the barycenter to a source.
$BC$ is a perpendicular from the point $B$ to the $xy$ plane, and $CD$
is a line parallel to $OB$; points $C$ and $D$ belong to the $xy$ plane.
The unit vector pointed from the barycenter to a source
has Cartesian coordinates
\hbox{$(\cos\alpha\cos\delta, ~\sin\alpha\cos\delta, ~\sin\delta)$.}
The scalar product of this vector and the radius-vector of the space
probe is the length of the segment $OB$.
$$
 OB = x_0 \cos\delta\cos\alpha + y_0 \cos\delta\sin\alpha +
      z_0 \sin\delta
$$
\noindent
A time interval during which
light flies the length $OB$ is a barycenter
correction, if a source is far enough and $SB$ is a part of a flat wave
front. This time interval $\Delta T$, which must be added to the moment
of each event in the point $S$ to obtain a corresponding barycenter
moment, is determined as
$$
 \Delta T = \frac{1}{c} \left( x_0 \cos\delta\cos\alpha +
                               y_0 \cos\delta\sin\alpha +
                               z_0 \sin\delta \right)   \eqno (1)
$$
\noindent
In a common choice $\alpha$ and $\delta$ coincide with
right ascension and declination, when $xy$ is an Earth's
equatorial plain for some fixed epoch and $x$-axis points to
spring equinox at the same epoch.
Procedure of calculation of $\Delta T$ with precise account of Earth
and satellite motion is described in \cite{cosbdbase}.

Suppose that coordinates $\alpha$ and $\delta$ are known not exactly
with corresponding errors $d\alpha$ and $d\delta$. Then for small
errors we may find from (1) corresponding barycenter corrections
in linear approximation

$$
\begin{array}{rcl}
 \delta T&=&{\displaystyle
             \frac{\partial\Delta T}{\partial\alpha} d\alpha +
             \frac{\partial\Delta T}{\partial\delta} d\delta =}\\
          &=&{\displaystyle
             \frac{1}{c} \Bigl(-x_0 \cos\delta\sin\alpha +
                                y_0 \cos\delta\cos\alpha\Bigr) d\alpha +\strut}
                                \rule{0pt}{6mm}                               \\
          &+&{\displaystyle
             \frac{1}{c} \Bigl(-x_0 \sin\delta\cos\alpha -
                                y_0 \sin\delta\sin\alpha +
                                z_0 \cos\delta\Bigr) d\delta}
                                \rule{0pt}{6mm}             \\
\end{array}
                                    \eqno (2)
$$
\noindent
    Consider for simplicity a case when a space probe orbit around
the Sun is circular and, consequently, its angular
velocity $\omega$ is a constant. Then
$x_0 = R\cos\omega t$, $y_0 = R\sin\omega t$, $z_0=0$ and
$$
  \Delta T = \frac{R}{c} (\cos\omega t \cos\delta \cos\alpha +
                          \sin\omega t \cos\delta \sin\alpha)
           = \frac{R}{c} \cos\delta \cos(\omega t - \alpha)
                          \eqno (3)
$$
\noindent
where $R$ is a radius of its orbit. The correction (2) then is
reduced to
$$
 \delta T =  \frac{R}{c} \Big[\cos\delta\sin(\omega t-\alpha ) \,d\alpha
             -             \sin\delta\cos(\omega t-\alpha) \,d\delta\Big].
                           \eqno (4)
$$
\noindent
Assume that only first and second derivatives of $\nu$ are essential, and
a frequency of the signal may be represented by

$$ \nu=\nu_0+\dot \nu_0 t+{\ddot \nu_0^2 \over 2}t^2,  \eqno(5)$$
\noindent
where index "0" is referred to the epoch $t=0$. Define a current arrival
time of photons from the source, measured on the satellite, as
$\tilde t$. When the source coordinates are known exactly, barycenter
arrival time $t$ is found as

$$t=\tilde t+\Delta T, \eqno(6)$$
\noindent
and having barycenter arrival times it is possible to find
$\nu_0$, $\dot \nu_0$ and $\ddot \nu_0$ using the criteria from
previous sections.

When source coordinates are not known exactly and their possible errors
are  $d\alpha$ and $d\delta$, the error in barycenter correction
is determined by (4). In order to estimate an input of these errors on
timing characteristics let us compare phases of the arriving signal
calculated from measurements $\phi'$ (with errors) and in
true barycenter time $\phi$, so that

$$ \phi'=\int_0^{t'} \nu' dt', \quad \phi=\int_0^{t} \nu dt \eqno(7) $$
\noindent
Here $t'$ is the time calculated from (6), and $\nu'$ is a frequency
found after barycenter corrections, containing errors. Times $t$ and
$t'$ correspond to the same event, so we may rewrite $\phi'$ in
true barycenter coordinates as

$$ \phi'=\int_0^t \biggl[1+{d\delta T \over dt}\biggr] dt \eqno(8) $$
\noindent
Using (4) and (5) in (8) we obtain after integration and account of
(4),(5)

$$ \phi'={\rm const}+\int_0^t \nu dt+\nu \delta T-
   \int_0^t \delta T (\dot \nu_0+\ddot\nu_0 t)dt $$
$$   ={\rm const}+
   \int_0^t \nu dt+\left(\nu-{\ddot \nu_0 \over \omega^2}\right)\delta T
   -(\dot \nu_0+\ddot\nu_0 t)\int_0^t \delta T dt. \eqno(9) $$
\noindent
Here and farther relations

$$\ddot{\delta T}=-\omega^2 \delta T, \quad
  \int(\int \delta Tdt)dt=-{\delta T \over \omega^2}, \quad
  {d\delta T \over dt}=-\omega^2 \int \delta T dt  \eqno(10)$$
\noindent
are used, and it follows from (4)

$$ \int \delta T dt=-{R \over c\omega}[\cos \delta \cos (\omega t-\alpha)
   d\alpha+\sin \delta \sin(\omega t-\alpha) d\delta]. \eqno(11)$$
\noindent
Differentiating (9) we obtain an input of the angular coordinate
errors into the values of frequency and its derivatives

$$ \nu'={d\phi' \over dt}=\nu\left(1+{d\delta T \over dt}\right),$$
$$\dot \nu'={d^2\phi' \over dt^2}=\dot\nu\left(1+{d\delta T \over dt}\right)
  -\nu \omega^2 \delta T, \eqno(12)$$
$$\ddot \nu'={d^3\phi' \over dt^3}=\ddot \nu+
(\ddot\nu-\nu \omega^2){d\delta T \over dt}
  -2\dot\nu \omega^2 \delta T, $$
\noindent
where

$$\dot \nu=\dot \nu_0+\ddot \nu_0 t, \quad \ddot \nu = \ddot \nu_0,
                     \eqno(13)$$
\noindent
and current values $\nu'$, $\dot \nu'$ and $\ddot \nu'$ are connected
with corresponding values at $t=0$ as

$$
\begin{array}{rcl}
 \nu'
   &=&{\displaystyle
       \nu'_0 + {\dot\nu}'_0 t +
      {\ddot\nu}'_0 \frac{t^2}{2}} \\
 \nu'_0
   &=&{\displaystyle
       \nu' - {\dot\nu}'t +
      {\ddot\nu}' \frac{t^2}{2}} \\
 {\dot\nu}'_0
   &=&{\dot\nu}' - {\ddot\nu}' t \\
 {\ddot\nu}'_0
   &=&{\ddot\nu}'
\end{array}
                        \eqno (14)
$$

   The detailed variant of previous calculations can also be
found in \cite{Prep1970}.

   In presence of a proper motion of the source the errors
$d\alpha$ and $d\delta$ change linearly in first approximation as

$$ d\alpha=d\alpha_0+ v_\alpha t, \quad
   d\delta=d\delta_0+ v_\delta t. \eqno(15)
$$
\noindent
It leads to farther complication of the formula (9)-(12). Note that in
simulations we deal not with these formulae, but directly with arrival
times of photons $\tilde t_i$, $t'_i$ and $t_i$. In the problem of timing
of radiopulsars the precision of observational data is very high, so
appearance of the periodical 1 year component gives a direct indication
to the errors in angular coordinates of the pulsar, possibility to
improve them \cite{taylor,harrison}, and to determine a proper motion.
In periodic sources with rare pulses a quality of data is much worse
and other methods, based on above mentioned criteria must be used.

\section{Mathematical simulation}
For checking a possibility to use criteria considered above for
determination of corrections to the angular coordinates and proper motion,
in addition to frequency and its two derivatives, artificial sample of
data was produced. A pulse shape was taken as $\delta$-function with
a frequency of the signal changing in time according to (5), what
corresponds to a phase dependence

$$ \phi=\phi_0+\nu_0 t+\dot \nu_0 \frac{t^2}{2}+
   \ddot \nu_0 \frac{t^3}{6}. \eqno(16)$$
\noindent
We need to find time moments, corresponding to phase values $\phi_i=2\pi i$.
Two sets of input parameters were considered. The time $t=0$ is related
to a point of the orbit, where $\alpha=0$.

$$
  ({\rm i})\,\,\,\phi_0=0,\,\, \nu_0=4\, {\rm s}^{-1},\,\,\dot\nu_0=
  -2\cdot10^{-8}\,{\rm s}^{-2},\,\,
  \ddot\nu_0=3\cdot10^{-16}\, {\rm s}^{-3},$$
$$   \omega=6.060171\cdot10^{-6}\, {\rm s}^{-1} \eqno(17)$$
$$ ({\rm ii})\,\,\, \phi_0=0,\,\, \nu_0=4\, {\rm s}^{-1},\,\,
   \dot\nu_0=-2\cdot10^{-13}\,
   {\rm s}^{-2},\,\, \ddot\nu_0=3\cdot10^{-26}\, {\rm s}^{-3},$$
$$\omega=1.991063802\cdot10^{-7}\, {\rm s}^{-1}$$
\noindent
     The source parameters are chosen to satisfy a relation
$\nu_0\ddot\nu_0/\dot\nu_0^2=3$, supposed to be valid for ejecting
pulsars \cite{TaylorManchester}.
Because of low reliability of $\ddot\nu$ detecting in gamma
observations, some authors \cite{Mattox1} set it equal to zero.
The modeling year duration
$(2\pi/\omega)$ is equal to 12 days in the first case; and is a
true value of 365.2422 days
in the second, when the moment $t=0$ corresponds to 21 March.
Note that in the second case the values of $\nu_0$ and $\dot\nu_0$
are chosen very close to that of Geminga \cite{circ5541}.

  One possible way to find $t_i$ is to use Burmann-Lagrange
expression, which links the Taylor coefficients of direct and
inverse functions.  We have used instead a procedure, valid
for a general law of a phase dependence $\phi(t)$, based on a Taylor
expansion formula
$$
  t = \sum_{n=0}^\infty \frac{1}{n!} \left(\frac{d^nt}{d\phi^n}\right)_0
      \phi_i^n=\sum_{n=0}^\infty \frac{A_n}{n!} \phi_i^n.    \eqno (18)
$$
\noindent
To find the coefficients $A_n=\left(\frac{d^nt}{d\phi^n}\right)_0$,
$n \ge 1$, use an evident equality

$$
    \dot\phi \frac{dt}{d\phi} = 1.        \eqno (19)
$$
\noindent
That gives

$$A_1=\frac{dt}{d\phi}_0=\frac{1}{\dot \phi_0}. \eqno(20)$$
\noindent
Differentiating (19) $(n-1)$ times over $t$ we obtain a relation, linear
to $A_n$, what permits to express $A_n$ as a function of $A_m$, $m \le n-1$,
and $\deriv{\phi}{m}$, $\deriv{\phi}{n}$ .
 As an example, after 5 differentiation we get

$$
  A_1\deriv{\phi_0}{6} +
  6A_2 \deriv{\phi_0}{5}\dot\phi_0 +
  15A_2 \deriv{\phi_0}{4}\ddot\phi_0 +
  10A_2 \tdots\phi_0^2 +
  15A_3 \deriv{\phi_0}{4}\dot\phi_0^2 +
  60A_3 \dot\phi_0\ddot\phi_0\tdots\phi_0 $$
 $$ +15A_3 \ddot\phi_0^3 +
  20A_4 \dot\phi_0^3\tdots\phi_0 +
  45A_4 \dot\phi_0^2 \ddot\phi_0^2 +
  15A_5 \dot\phi_0^4 \ddot\phi_0 +
    A_6 \dot\phi_0^6=0,   \eqno(21)
$$
\noindent
where index $"0"$ indicates time $t=0$.
For (16) with $\,\,\deriv{\phi_0}{m}=0$
at $m \ge 4$ we have $\dot\phi_0=\nu_0$,
$\ddot\phi_0=\dot \nu_0$, $\tdots\phi_0 = \ddot \nu_0$ and get from (21)
an equation for $A_6$

$$
  10A_2 \ddot\nu_0^2 +
  60A_3 \nu_0\dot\nu_0\ddot\nu_0 +
  15A_3 \dot\nu_0^3 +
  20A_4 \nu_0^3\ddot\nu_0 +
  45A_4 \nu_0^2 \dot\nu_0^2 +
  15A_5 \nu_0^4 \dot\nu_0 +
    A_6 \nu_0^6=0   \eqno(22)
$$
\noindent
     The first criterion $K_1$ of periodicity
\cite{Gur1,Gur5,Gur4,Gur6}
was used to investigate periodicity
properties of series of pulses. For a purpose of testing
short intervals of ``observation'' were taken
in a different parts of the year. The error in coordinates was taken
equal to 5 arc seconds in absolute value $\sqrt{d\alpha^2+d\delta^2}$,
but the deviations from the initial point were taken in
eight different
directions, separated by $45^\circ$. The values of
$\nu'_0,\, {\dot\nu_0'},\, {\ddot\nu_0'}$ that have been
detected by $K_1$ criterion coincide in both cases with a very high
precision with theoretical ones from (12)-(14), see Table 1.
It may be seen from Table 1 a strong influence  of the errors
on the determination of $\ddot \nu'$ by using a criterion.
While the error in $\ddot \nu'$ is almost linearly proportional to
the error in $\sqrt{d\alpha^2+d\delta^2}$, it is evident that at
an angular error larger then $10^{-3}$ arc sec direct determination of
$\ddot \nu'$ by criterion becomes impossible. This may be a reason
for a high breaking index of Geminga \cite{circ5541,bispropmov}.

Let us now formulate a problem of timing of a gamma pulsar, which gives a
possibility for a search of its timing properties together with
angular coordinates and a proper motion. Assume that a gamma pulsar
simulated by computer
is emitting signals with a frequency changing according to (5),
satisfying condition $\nu_0 \ddot \nu_0/\dot \nu_0^2=3$.

Let the signal
registered on the probe is reduced to the barycenter time, using
the source coordinates $\alpha_0$ and $\delta_0$ (base point),
which contain errors $d\alpha_0$ and $d\delta_0$
respectively. We suspect also a proper motion of the source
defined by following parameters: at the
moment $t_0=0$ the source has coordinates $\alpha_0 + d\alpha_0$ and
$\delta_0 + d\delta_0$ and a velocity modulus $\dot \theta={\rm const}$.
A velocity direction is defined by an angle $\theta$, which
is counted clockwise from the positive $\alpha$-axis direction.
The current source coordinates are consequently
$$
 \alpha = \alpha_0 + d\alpha_0 + v_{_<} t\cos\theta,  \qquad
 \delta = \delta_0 + d\delta_0 + v_{_<} t\sin\theta,  \eqno (23)
$$

     Thus, we have seven parameters that are needed
to be found selfconsistently

\begin{tabular}{ccp{12cm}}
 $\nu_0$        & -- & frequency of the source signal        \\
 ${\dot\nu_0}$  & -- & first derivative of the frequency     \\
 ${\ddot\nu_0}$ & -- & second derivative of the frequency    \\
 $d\alpha_0$          & -- & shift in right ascension from the
                             base point $(\alpha_0, \delta_0)$ at
                             $t_0$ epoch                           \\
 $d\delta_0$
                      & -- & shift in declination from the base
                             point $(\alpha_0, \delta_0)$ at $t_0$
			     epoch                                 \\
 $v_{_<}$             & -- & proper angular velocity of
			     the source in celestial coordinates  \\
 $\theta$             & -- & direction of velocity $v_{_<}$,
                             counted clockwise from the positive
                             $\alpha$-axis direction \\
 \end{tabular}

 \noindent\looseness=-1
     with additional restriction:
$\nu_0\ddot\nu_0/\dot\nu_0^2 = 3$ following from the model of the
pulsar radiation \cite{TaylorManchester}.

     In data simulation we fix parameters
$\nu_0, \dot\nu_0, \ddot\nu_0$ as (ii) in (17), find true
$\alpha$ and $\delta$  from (23) with
$d\alpha_0 = 2^{\prime\prime}$, $d\delta_0 = 3^{\prime\prime}$,
$v_{_<} = 0^{\prime\prime}\!\!.2$ per year, $\theta = 60^\circ$,
and create the
simulated data, i.e. the sequence of time moments of pulses.
Time moments (true barycenter) found for a source from (19)-(22)
are then recalculated for a probe using correct coordinates and (23).
Now to a set of time moments "registered" by a probe from the source
with subscribed
coordinates $\alpha_0$ and $\delta_0$, containing errors,  we
apply the algorithm for searching the periodic signal to extract
all the seven parameters from the simulated data set.
Namely, we consider a number of different sets of 7 mentioned
parameters. For each set we evaluate supposed errors,
introduced in the data due to errors in a position and in a
proper motion of the object. After that we subtract these supposed
errors from the data.
Then, assuming the data free of errors the periodicity
criterion value was calculated. Remind, that if we assume the data
free of errors, it means that the phases of the pulses must obey
the simple relation (16).
The criterion reaches its absolute maximum only for an exact set
of parameters.
There appear
a number of local (or false) maxima. Fig.~5
demonstrates
a typical structure of the criterion depending on two parameters:
$\nu$ and ${\dot\nu}$, when other five ones are fixed
(see also \cite{Mattox1}).
Because the height
of a ``false'' maxima is close to ``true'' one, it is very
difficult to separate the absolute maximum among a series of local
ones.

\looseness=-1
We are looking for an absolute maximum,
using a grid in 4-dimensional space \linebreak
$(d\alpha_0, d\delta_0, v_{_<}, \theta )$, that
was defined by the following way:
   $d\alpha_0$ varies from $-5^{\prime\prime}\,\,$
                         step $0^{\prime\prime}\!\!.5$
                      to   $5^{\prime\prime}\,\,$,
   $d\delta_0$ varies from $-5^{\prime\prime}\,\,$
                      step $0^{\prime\prime}\!\!.5$
                      to   $5^{\prime\prime}\,\,$,
   $v_{_<}$    varies from $0^{\prime\prime} /$year
                      step $0^{\prime\prime}\!\!.05 /$year
                      to   $0^{\prime\prime}\!\!.5 /$year and
   $\theta$ varies    from $0^\circ\,\,$
                      step $10^\circ\,\,$
                      to   $350^\circ\,\,$.
Three other parameters
($\nu_0,\, {\dot\nu_0},\, {\ddot\nu_0}\,$) were detected jointly
for each point of above grid,
using the grid $6\times6\times6$ with 12 times
consecutively diminishing steps for all three axes. The maximum,
detected at the previous step was placed to the center of the
grid and all the scale multiplied by the factor of 0.4. This
procedure was repeated for 12 times, so that the last grid steps
are $0.4^{12}\approx 1.7\cdot 10^{-5}$ times as small as
the first ones.
All seven parameters chosen for modeling were found with a precision,
limited only by computational grid connected with a power of the computer,
using criterion $K_1$.
So, for a clean set of data the proposed procedure
of searching is working effectively. Situation is becoming much
more controversial when we apply it to real data existing to the
moment.

\looseness=0

\section{Application to Geminga}
  \subsection{Analysis of COS--B data}

     COS--B mission had been operated since August, 1975 till April 1982,
and had observed Geminga
in five shifts \cite{cosbdbase,bignami1}.
Their numbers are 00, 14, 39, 54 and 64. Some measurements of Geminga
position, reduced to the epoch 1950.0, are summarized in
Table 2 and are plotted in Fig.~6.

     It was obtained in \cite{circ5541} from
COS--B data  $\nu=4.217$ Hz, $\dot \nu=-1.952
\cdot 10^{-13}$Hz$\cdot$s$^{-1}$, and a large value of
$\ddot\nu = (28\pm16)\cdot10^{-26}$ Hz$\cdot$s$^{-2}$, corresponding to
a braking index $n=31 \pm 18$. Barycenter corrections have
been done for standing Geminga with coordinates No. 3,8 in Table 2.
Quanta selection used in our analysis has been done by two different ways:

\begin{description}
 \parshape=1 2cm 14cm
 \item[1)]
   all the quanta in the circle of $r=5^\circ$ around Geminga
   position; the interval 54 was excluded because of low reliability;
   it is total of 1505 quanta.
 \item[2)]
   all the quanta of the energy $E > 50$ MeV laying in the circle
   $r=12.5\cdot E^{-0.16}$, where $E$ is measured in MeV and $r$
   in degrees \cite{buccheri2}; total of 1883 quanta.
\end{description}

\noindent
The second selection is close to that used in \cite{bignami1}.
A problem of a quanta selection criterion is a very delicate one
because it is practically impossible for a single quantum to decide,
was it really radiated by Geminga or belongs to a background.
Both mentioned selections are
noisy, but the first one is worse.

     Geminga was considered as a moving object. The base coordinates,
that are used for initial barycenter reduction are the position
measured by Einstein satellite in 1981 (see Table~2).
Geminga motion was defined by its velocity, direction and
initial position at the epoch 1979, March, 14.0 \cite{circ5541}.
The obtained barycenter time moments for each quantum was
additionally reduced to the barycenter, using expression (2).
A motion of the probe defined by $x_0(t),\,y_0(t),\,z_0(t)$ was taken
from databases of COS B \cite{cosbdbase} or EGRET.
The object coordinates were calculated by this procedure separately
for each quantum according to a supposed object motion.

The results, with using the second selection from the mentioned
above, are not very certain. The criterion appear to
exhibit a gently sloping maximum at the following model parameters:
velocity
$v_{_<} = 0^{\prime\prime}\!\!.2 - 0^{\prime\prime}\!\!.3$ per year,
direction $\theta = 40^\circ - 60^\circ$ and the initial
coordinate offsets $d\alpha_0$ and $d\delta_0$ at the mentioned
epoch are $-2^{\prime\prime}$ for
both $\alpha$- and $\delta$-axes, but the uncertainty here is
high and may reach $2^{\prime\prime}$ for both coordinates.
As to the periodicity parameters, they are in a good agreement
with \cite{circ5541}, except the second derivative ${\ddot\nu_0}$,
which is a bit smaller but lays within the error box of standing
Geminga. The motion of Geminga, obtained in our investigation
does not contradict to the motion of G$^{\prime\prime}$ star \cite{bignami4}.

     Unfortunately, this solution is not a unique one, and
there are a number of other maxima of approximately
the same height. When we
search for a global maximum in 7-dimensional space
it is extra difficult to detect ``the main maximum'' among a series
of other local maxima. For example, there is an accessory maximum
at $v_{_<} = 0^{\prime\prime}\!\!.1 - 0^{\prime\prime}\!\!.2$ per year,
$\theta = 340^\circ - 360^\circ$ and very badly detected initial
offsets (it can only be said that they both are negative). The
periodicity parameters here are approximately the same as above,
but ${\ddot\nu_0}$ appears to be negative.

\subsection{Analysis of EGRET data}

     EGRET experiment is operating since April 1988. There are 9
periods of observations where Geminga was not far from the center
of a view field (less than $30^\circ$, the standard requirement).
The following sessions was
used for data investigations: 2, 3, 4, 5, 10, 21, 2130, 2210, 3100.
     The standard procedure for the barycenter correction
was used \cite{FITS,FITSCook},
but we have used coordinates of the object from \cite{gempos}, line 8 in
Table 2, different from
those, indicated in EGRET data base. We have used the same Geminga
coordinates for both satellites, COS-B and EGRET. They are the best
fit Einstein position, but in the last
case we were to reduce them to the epoch 2000.0, because it is used in an
appropriate barycenter reduction routines.  Reduced coordinates
are the following:

$$
   \alpha_{2000} = 98^\circ 28^\prime 30^{\prime\prime}\!\!.90
          = 98^\circ\!\!.47525 \qquad\qquad
   \delta_{2000} =+17^\circ 46^\prime 11^{\prime\prime}\!\!.6
          =+17^\circ\!\!.76989       \eqno (24)
$$

     And in EGRET data base the coordinates are:

$$
   \alpha_{2000}= 98^\circ\!\!.48
          = 98^\circ 28^\prime 48^{\prime\prime}\!\!\qquad\qquad
   \delta_{2000}=+17^\circ\!\!.77
          =+17^\circ 46^\prime 12^{\prime\prime}\!\!
$$

     Other authors \cite{CaravCoord,Mattox1} use the coordinates
close to (24), they differ less than $1^{\prime\prime}$ from
the center of Einstein error box (lines 3, 8 in the Table~2).

     We have used a number of techniques for quanta selection
and have compared the results. The selections used are the following:

\begin{description}
 \parshape=1 2cm 14cm
 \item[1)]
   all the quanta of the energy $E > 70$ MeV laying in the circle
   $r=5.85\cdot (E/100)^{-0.534}$, where $E$ is measured in MeV and $r$
   in degrees \cite{thompson3}; it is so-called the standard selection;
   total of 6751 quanta.
 \item[2)]
   all the quanta of the energy $E > 1500$ MeV in the circle
   $r = 2^\circ$; total of 365 quanta.
 \item[3)]
   all the quanta of the energy $E > 2000$ MeV in the circle
   $r = 2^\circ$; total of 223 quanta.
 \item[4)]
   all the quanta of the energy $E > 2000$ MeV in the circle
   $r=0^\circ\!\!.5$
   round the Geminga position; total of 100 quanta.
 \item[5)]
   all the quanta of the energy $E > 3000$ MeV in the circle
   $r=0^\circ\!\!.5$
   round the Geminga position; total of only 51 quanta.
\end{description}

     The criterion value
for unmoving Geminga, resting in Einstein's HRI position is
$K_1 = 0.0572$ and the parameters of periodicity
are $\nu_0^{\prime}=4.21775012925$ Hz,
${\dot\nu_0}^{\prime}=-1.95312\cdot 10^{-13}$ Hz$\cdot$s$^{-1}$,
${\ddot\nu_0}^{\prime}=(20\pm12) \cdot 10^{-26}$ Hz$\cdot$s$^{-2}$.
An appropriate light curve is shown in Fig.~7.
The results of search in 7-dimensional space
for different selections are:

\begin{description}
 \item[1)]
  There is a gently sloping maximum in criterion value at
  the following parameters:
  $\nu_0=4.2177501295$ Hz,
  ${\dot\nu_0}=-1.9532\cdot 10^{-13}$ Hz$\cdot$s$^{-1}$,
  ${\ddot\nu_0}=(22 \pm 13)\cdot 10^{-26}$ Hz$\cdot$s$^{-2}$,
  $v_{_<}=0^{\prime\prime}\!\!.3 - 0^{\prime\prime}\!\!.4$ per year,
  $\theta=40^\circ - 60^\circ$,
  $d\alpha_0$ and $d\delta_0$ are defined with very low
  precision and both lay in the interval $-2^{\prime\prime}$
  to $0^{\prime\prime}$.
  The criterion value $K_1=0.05732$.
  The second derivative $\ddot\nu$ here is of rather large value,
  so that the braking index is approximately equal to 25.

 \item[2)] \tolerance=1000
  There is a gentle maximum in criterion value at
  the following parameters:
  $\nu_0=$ 4.2177501273 Hz,
  ${\dot\nu_0}=-1.952711\cdot 10^{-13}$ Hz$\cdot$s$^{-1}$,
  ${\ddot\nu_0}\approx (0\pm 10)\cdot 10^{-26}$ Hz$\cdot$s$^{-2}$
  $v_{_<}=0^{\prime\prime}\!\!.3$ per year or more,
  $\theta=40^\circ - 60^\circ$,
  $d\alpha_0=0^{\prime\prime} -( -2^{\prime\prime}), \,\,
   d\delta_0=0^{\prime\prime} -( -1^{\prime\prime})$
  and criterion value $K_1=0.22357$.
  The second derivative here is very close to zero and
  the braking index is, respectively, also low and could be close
  to its theoretical value.

 \item[3)] \tolerance=400
  There is a gentle maximum in criterion value
  at approximately the following parameters:
  $\nu_0=4.2177501261$ Hz,
  ${\dot\nu_0}=-1.952611\cdot 10^{-13}$ Hz$\cdot$s$^{-1}$,
  ${\ddot\nu_0}=(-2.4\pm 10)\cdot 10^{-26}$ Hz$\cdot$s$^{-2}$,
  $v_{_<}=0^{\prime\prime}\!\!.3 - 0^{\prime\prime}\!\!.4$ per year,
  $\theta=40^\circ - 80^\circ$,
  $d\alpha_0=0^{\prime\prime} - (-2^{\prime\prime}),
   d\delta_0=-1^{\prime\prime} -( -3^{\prime\prime})$
  and the criterion value $K_1=0.25535$.
  It was impossible to determine the parameters with higher precision.

 \item[4)] \tolerance=1000
  There is a gentle maximum in criterion value
  at the following parameters:
  $\nu_0=$ 4.2177501344 Hz,
  ${\dot\nu_0}=-1.954477\cdot 10^{-13}$ Hz$\cdot$s$^{-1}$,
  ${\ddot\nu_0}= (68\pm 40)\cdot 10^{-26}$ Hz$\cdot$s$^{-2}$,
  $v_{_<}=0^{\prime\prime}\!\!.3 - 0^{\prime\prime}\!\!.4$ per year or more,
  $\theta=220^\circ - 260^\circ$,
  $d\alpha_0$ and $d\delta_0$ are negative
  and the criterion value $K_1=0.3072$.
  Because of a very poor statistics this selection
  as well as the following one can be considered as a test only.
  The line of motion here is the same as in the previous
  cases, but the direction is opposite .

 \item[5)] \tolerance=200
  There is a relatively good maximum despite of a very poor
  statistics at the following parameters:
  $\nu_0=4.2177501220$ Hz,
  ${\dot\nu_0}=-1.95282\cdot 10^{-13}$ Hz$\cdot$s$^{-1}$,
  ${\ddot\nu_0}=(12\pm 10)\cdot 10^{-26}$ Hz$\cdot$s$^{-2}$,
  $v_{_<}=0^{\prime\prime}\!\!.2 - 0^{\prime\prime}\!\!.3$ per year,
  $\theta=20^\circ - 60^\circ$,
  $d\alpha_0=0^{\prime\prime} - -2^{\prime\prime},
   d\delta_0 \approx 0^{\prime\prime}$
  and the criterion value $K_1=0.44441$.
  The reliability of this result is not high, but large
  criterion value indicates that we may deal with real motion
  of the object.

\end{description}

  \subsection{Analysis of combined COS-B and EGRET data}

   We have combined COS-B and EGRET data with the following selection
criteria:

\begin{description}
   \item[COS-B:] $E > 50$ MeV and the standard conditions for $r$:
		 $r=12.5\cdot E^{-0.16}$ ~~\cite{buccheri2}
   \item[EGRET:] $E > 70$ MeV and the standard conditions for $r$:
		 $r=5.85\cdot (E/100)^{-0.534}$ ~~\cite{thompson3}
\end{description}

     There are total of 1883 + 6751 = 8634 quanta.

     The combined series is not self-contradicting. There is
a gentle maximum in criterion value at the following parameters:
  $\nu_0=4.21775012323 \pm 0.000 000 00025$ Hz,
  ${\dot\nu_0}=(-1.952554 \pm 0.000 025)\cdot 10^{-13}$ Hz s$^{-1}$,
  ${\ddot\nu_0}=(-2.5 \pm 10)\cdot 10^{-26}$ Hz s$^{-2}$,
  $v_{_<}=0^{\prime\prime}\!\!.5-0^{\prime\prime}\!\!.6 $ per year,
  $\theta=55^\circ - 65^\circ$,
  $d\alpha_0=-2^{\prime\prime}-(-4^{\prime\prime}),
   d\delta_0=1^{\prime\prime}-2^{\prime\prime} $
  and the criterion value $K_1=0.04937$.

     Note that $1\sigma$ errors above were obtained by
approximate estimations.

\section{Discussion}

  According to our investigations the coordinates and
motion of Geminga obtained from timing of gamma pulsar
is in a satisfactory agreement with the
motion of $G^{\prime\prime}$ star when separately COS-B or EGRET
data are used. Parameters following from the combined data set are
in much worse agreement. There are two possible reasons of it.
First there could exist a systematic error between the data
of two probes; and second, period could behave nonmonotonously
between 1982 and 1988 years, and the period jump (pulsar glitch)
of the order of ${\Delta P \over P} \ge 10^{-10}$ could already
spoil the parameters obtained by criteria.

  The value of the second derivative ${\ddot\nu_0}$ does not
coincide with the theoretical one, however it is
lower than in the previous investigations, but there are weighty
reasons to explain this phenomenon.
The second derivative is a very sensitive variable,
and even $0^{\prime\prime}\!\!.001$ error in angular coordinates
changes ${\ddot\nu_0}$ by the value, comparable with the
result (see Table 1).
If there was a jump in pulsar period,
it may also cause the incorrect value of a variable.
Possibility to improve angular resolution by timing is strongly limited
by small number of quanta and existence of considerable background.

As a result of application to Geminga of the developed method
of timing of gamma pulsars we have obtained that determination of true
value of ${\ddot\nu_0}$ is possible only at very high precision
(better then $0^{\prime\prime}\!\!.001$) of angular localization.
At good statistics of gamma pulsars corresponding improvements
would become possible from timing analysis. New types of gamma ray
telescopes based on very wide aperture ($\ge 2.5\, \pi$ steradian) and
higher threshold of a few hundred MeV \cite{bkl,lbk} would permit
to get higher angular resolution ($\sim \,1$ arc min), reducing
influence of a background, get $\sim\, 100$ better statistics due to
continuous monitoring of larger part of the sky in this region.

For the existing data of Geminga from COS B and EGRET it was
obtained, using only gamma-ray data, that criterion value reaches its
maximum at nonzero value of a proper motion. The coordinates of
the source were confirmed with precision $\sim 2^{\prime\prime}\!\!$
what is better then follows from X ray data, but, of course, is
worse then the precision obtained in the optical observations.

Observations of radio pulsars have shown, that their optical and X-ray
luminosity is decreasing with time much more rapid, that radio and
hard gamma radiation. So at increasing sensitivity we expect
a discovery of tens of new gamma-ray pulsars similar to Geminga,
may be without
X-ray and optical counterparts. For such objects method of
timing of gamma-ray pulsars developed above would be a main and may be
a single means of investigation of such sources by data processing.

\section{Acknowledgement}
    The authors very grateful to Dr. D.Macomb for his help
in access to EGRET data base, and to Dr. P.Caraveo and
Dr. P.Goldoni for their help in choice of quanta selection criteria.

    This work was partly supported by Russian Foundation
for Base Research, grants No.~93-02-16553, 96-02-17231;
Program Astronomy, grant No.1.2.6.5 and CRDF award \# Rp1--173.

One author (S.V.R.) is very grateful to Prof. E.Starostenko,
Dr. A.Salpagarov and Dr. O.Sumenkova for their great help.

\large
\newpage
\centerline{
 {\bf Table 1.}}

\qquad\qquad\qquad Properties of a source with angular coordinates and

\qquad\qquad\qquad timing characteristics close to Geminga from (17),
case (ii),

\qquad\qquad\qquad calculated using criterion (1). Barycenter
corrections

\qquad\qquad\qquad have been done using angular coordinates with
indicated errors.
\hfill\break
\hfill\break
\hfill\break
\hfill\break
\centerline{
\begin{tabular}{|c|c|c|c|c|r|}
 \hline
  \multicolumn{6}{|c|}{\rule{0pt}{9pt}Source coordinates: \quad
                       $\alpha=6^h30^m00^s,\quad
                       \delta=+17^\circ30^\prime00^{\prime\prime}$}\\
 \hline\rule{0pt}{9pt}
  Month, & \multicolumn{2}{|c|}{Errors} &
          \multicolumn{3}{|c|}{Theoretical values} \\
 \cline{2-6}
  day    &$d\alpha, ^{\prime\prime}$&$d\delta, ^{\prime\prime}$&
          $\tilde\nu_0$, s$^{-1}$
          &\rule{0pt}{12pt}$\tilde{\dot\nu_0}, 10^{-13}$ s$^{-2}$
          &$\tilde{\ddot\nu_0}, 10^{-26}$ s$^{-3}$\\
 \hline
   &&&&&\\
   May, 20 &\phantom{$-$}5.00&\phantom{$-$}0.00& 3.99999999748 &$-$1.9738&$-28600$~~~~\\
   May, 20 &\phantom{$-$}2.24&\phantom{$-$}2.24& 3.99999999744 &$-$1.9879&$ -9620$~~~~\\
   May, 20 &\phantom{$-$}0.00&\phantom{$-$}5.00& 3.99999999681 &$-$1.9992&$  7140$~~~~\\
   May, 20 &          $-$2.24&\phantom{$-$}2.24& 3.99999999970 &$-$2.0114&$ 16000$~~~~\\
   May, 20 &          $-$5.00&\phantom{$-$}0.00& 4.00000000252 &$-$2.0262&$ 28600$~~~~\\
   May, 20 &          $-$2.24&          $-$2.24& 4.00000000256 &$-$2.0121&$  9620$~~~~\\
   May, 20 &\phantom{$-$}0.00&          $-$5.00& 4.00000000319 &$-$2.0008&$ -7140$~~~~\\
   May, 20 &\phantom{$-$}2.24&          $-$2.24& 4.00000000030 &$-$1.9989&$-16000$~~~~\\
   July, 10&\phantom{$-$}5.00&\phantom{$-$}0.00& 3.99999999621 &$-$1.9695&$-35700$~~~~\\
   Aug., 20&\phantom{$-$}5.00&\phantom{$-$}0.00& 4.00000000544 &$-$1.9852&$-22300$~~~~\\
   Aug., 20&\phantom{$-$}0.00&          $-$5.00& 4.00000001020 &$-$2.0155&$  9110$~~~~\\
   Nov., 20&\phantom{$-$}5.00&\phantom{$-$}0.00& 4.00000008067 &$-$2.0725&$ 29150$~~~~\\
   Nov., 20&          $-$2.24&\phantom{$-$}2.24& 3.99999996211 &$-$1.9631&$-16200$~~~~\\
   &&&&&\\
 \hline
\end{tabular}
}

\newpage

\centerline{
 {\bf Table 2.} Geminga position from different measurements
}
\hfill\break
\hfill\break
\centerline{
\begin{tabular}{|c|c|c|l|l|}
 \hline
 \multicolumn{5}{|c|}{Geminga position, epoch 1950}     \\
 \hline
 No. & $\alpha$  &  $\delta$  & Error &\qquad\qquad Comments  \\
 \hline                                          &&&&   \\
    1&
    $ 97^\circ44^\prime43^{\prime\prime}\!\!.7$  &
    $+17^\circ48^\prime27^{\prime\prime}\!\!.5$  &
    $12^{\prime\prime}$                          &
    ROSAT PSPC, 1991, Sep. 19-21, \cite{becker1}        \\
    2&
    $ 97^\circ44^\prime51^{\prime\prime}\!\!.7$  &
    $+17^\circ48^\prime36^{\prime\prime}\!\!.0$  &
    $\approx 5^{\prime\prime}$                   &
    ROSAT HRI, 1991, Mar. 19, \cite{becker1}            \\
    3&
    $ 97^\circ44^\prime47^{\prime\prime}\!\!.2$  &
    $+17^\circ48^\prime33^{\prime\prime}\!\!.0$  &
    $3^{\prime\prime}\!\!.2$                     &
    Einstein, 1981, Mar. 18, \cite{becker1}             \\
    4&
    $ 97^\circ44^\prime45^{\prime\prime}\!\!.9$  &
    $+17^\circ48^\prime32^{\prime\prime}\!\!.7$  &
    $0^{\prime\prime}\!\!.46$                    &
    G$^{\prime\prime}$ star, 1984, \cite{bignami4}      \\
    5&
    $ 97^\circ44^\prime45^{\prime\prime}\!\!.9$  &
    $+17^\circ48^\prime32^{\prime\prime}\!\!.6$  &
    $0^{\prime\prime}\!\!.5$                     &
    G$^{\prime\prime}$ star, 1986, Feb. 3, \cite{becker1}
                                                        \\
    6&
    $ 97^\circ44^\prime46^{\prime\prime}\!\!.5$  &
    $+17^\circ48^\prime33^{\prime\prime}\!\!.0$  &
    $0^{\prime\prime}\!\!.68$                    &
    G$^{\prime\prime}$ star, 1987, \cite{bignami4}      \\
    7&
    $ 97^\circ44^\prime47^{\prime\prime}\!\!.2$  &
    $+17^\circ48^\prime33^{\prime\prime}\!\!.6$  &
    $0^{\prime\prime}\!\!.16$                    &
    G$^{\prime\prime}$ star, 1992, \cite{bignami4}      \\
    8&
    $ 97^\circ44^\prime47^{\prime\prime}\!\!.2$  &
    $+17^\circ48^\prime33^{\prime\prime}\!\!.0$  &
    $3^{\prime\prime}\!\!.0$                     &
    Einstein, 1981, Mar. 18 \cite{gempos}               \\
                                                 &&&&   \\
 \hline
\end{tabular}
}

\newpage

\setcounter{figure}{3}

\begin{figure*}[h]
\rule{35mm}{0pt}
 \begin{picture}(160,100)
  \thicklines
  \put(0,0){\line(1,0){160}}
  \put(160,0){\line(0,1){100}}
  \put(160,100){\line(-1,0){160}}
  \put(0,100){\line(0,-1){100}}
  \multiput(0,0)(0,20){5}{\line(-1,0){2}}
  \multiput(16,0)(16,0){9}{\line(0,1){2}}
  \put(-22,-2){\phantom{00}0}
  \put(-22,18){100}
  \put(-22,38){200}
  \put(-22,58){300}
  \put(-22,78){400}
  \put(-35,90){Pulses}
  \put(66,4){Phase}
  \put(170,90){Epoch 1979, March, 14.0}
  \put(170,76){$\tilde\nu_0 = 4.21775012925465$ Hz}
  \put(170,62){$\tilde{\dot\nu_0} = -1.95316166\cdot 10^{-13}$ Hz s$^{-1}$}
  \put(170,48){$\tilde{\ddot\nu_0} = 19.66\cdot 10^{-26}$ Hz s$^{-2}$}
  \put(170,34)
    {$n = \tilde\nu_0\tilde{\ddot\nu_0}/\tilde{\dot\nu_0}^2 = 21.73$}
  \put(0,20){\line(1,0){4}}
  \put(4,20){\line(0,1){3}}
  \put(4,23){\line(1,0){4}}
  \put(8,23){\line(0,-1){5}}
  \put(8,18){\line(1,0){4}}
  \put(12,18){\line(0,1){0}}
  \put(12,18){\line(1,0){4}}
  \put(16,18){\line(0,-1){1}}
  \put(16,17){\line(1,0){4}}
  \put(20,17){\line(0,1){4}}
  \put(20,21){\line(1,0){4}}
  \put(24,21){\line(0,1){2}}
  \put(24,23){\line(1,0){4}}
  \put(28,23){\line(0,1){3}}
  \put(28,26){\line(1,0){4}}
  \put(32,26){\line(0,1){9}}
  \put(32,35){\line(1,0){4}}
  \put(36,35){\line(0,1){18}}
  \put(36,53){\line(1,0){4}}
  \put(40,53){\line(0,1){19}}
  \put(40,72){\line(1,0){4}}
  \put(44,72){\line(0,-1){8}}
  \put(44,64){\line(1,0){4}}
  \put(48,64){\line(0,-1){17}}
  \put(48,47){\line(1,0){4}}
  \put(52,47){\line(0,-1){9}}
  \put(52,38){\line(1,0){4}}
  \put(56,38){\line(0,-1){6}}
  \put(56,32){\line(1,0){4}}
  \put(60,32){\line(0,-1){3}}
  \put(60,29){\line(1,0){4}}
  \put(64,29){\line(0,1){2}}
  \put(64,31){\line(1,0){4}}
  \put(68,31){\line(0,-1){3}}
  \put(68,28){\line(1,0){4}}
  \put(72,28){\line(0,1){1}}
  \put(72,29){\line(1,0){4}}
  \put(76,29){\line(0,-1){1}}
  \put(76,28){\line(1,0){4}}
  \put(80,28){\line(0,1){3}}
  \put(80,31){\line(1,0){4}}
  \put(84,31){\line(0,1){2}}
  \put(84,33){\line(1,0){4}}
  \put(88,33){\line(0,-1){3}}
  \put(88,30){\line(1,0){4}}
  \put(92,30){\line(0,1){1}}
  \put(92,31){\line(1,0){4}}
  \put(96,31){\line(0,1){0}}
  \put(96,31){\line(1,0){4}}
  \put(100,31){\line(0,1){0}}
  \put(100,31){\line(1,0){4}}
  \put(104,31){\line(0,1){5}}
  \put(104,36){\line(1,0){4}}
  \put(108,36){\line(0,1){0}}
  \put(108,36){\line(1,0){4}}
  \put(112,36){\line(0,1){3}}
  \put(112,39){\line(1,0){4}}
  \put(116,39){\line(0,1){24}}
  \put(116,63){\line(1,0){4}}
  \put(120,63){\line(0,1){26}}
  \put(120,89){\line(1,0){4}}
  \put(124,89){\line(0,-1){20}}
  \put(124,69){\line(1,0){4}}
  \put(128,69){\line(0,-1){31}}
  \put(128,38){\line(1,0){4}}
  \put(132,38){\line(0,-1){12}}
  \put(132,26){\line(1,0){4}}
  \put(136,26){\line(0,-1){2}}
  \put(136,24){\line(1,0){4}}
  \put(140,24){\line(0,-1){4}}
  \put(140,20){\line(1,0){4}}
  \put(144,20){\line(0,-1){1}}
  \put(144,19){\line(1,0){4}}
  \put(148,19){\line(0,1){3}}
  \put(148,22){\line(1,0){4}}
  \put(152,22){\line(0,1){0}}
  \put(152,22){\line(1,0){4}}
  \put(156,22){\line(0,-1){8}}
  \put(156,14){\line(1,0){4}}
 \end{picture}
\hfill\break
\hfill\break
\hfill\break
\caption{
Geminga 40-bin light curve from ERGET data,
for standing Geminga with cooordinates No.~8 from Table 2
and standard quanta selection}
\label{GemLc1}
\end{figure*}
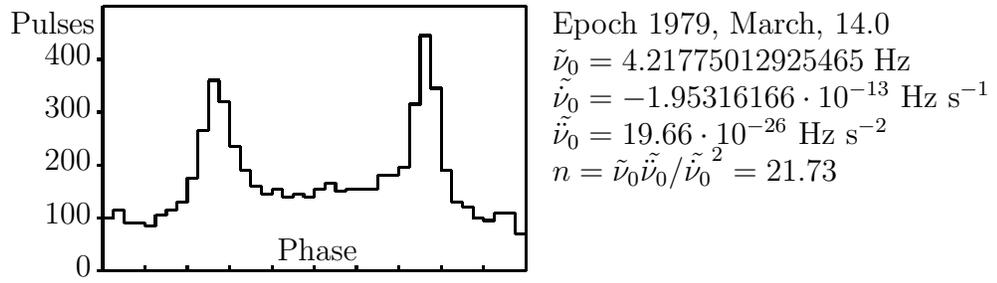

\newpage

\bigskip\bigskip
{\centerline{\bf Figure captions}}

\bigskip\bigskip
Figure 1: On the barycentric correction.

\bigskip\bigskip
Figure 2: On the absolute maximum structure.

\bigskip\bigskip
Figure 3: On the Geminga position.

\bigskip\bigskip
Figure 4: Geminga 40-bin light curve from ERGET data,
for standing Geminga with

$\qquad\quad$ cooordinates No.~8 from Table 2
and standard quanta selection.


\newpage

\setcounter{figure}{0}

\begin{figure*}[h]
\rule{50mm}{0pt}
 \begin{picture}(190,100)
 \linethickness{0.6mm}
 \put(0,0){\line(1,0){120}}
 \thicklines
 \put(120,0){\line(1,1){60}}
 \put(180,60){\line(-1,0){120}}
 \put(60,60){\line(-1,-1){60}}
 \thinlines
 \put(90,30){\circle*{3}}
 \put(90,22){\scriptsize O}
 \put(30,30){\line(1,0){120}}
 \put(90,30){\vector(1,0){70}}
 \put(162,30){\small$x$}
 \put(60, 0){\vector(1,1){70}}
 \put(132,70){\small$y$}
 \put(90,30){\vector(0,1){45}}
 \put(92,75){\small$z$}
 \put(90,30){\vector(4,3){60}}
 \put(50,0){\line(-4,-3){10}}
 \put(144,77){\scriptsize source}
 \put(140,42){\circle*{2}}
 \put(143,40){\scriptsize S}
 \put(136,39){\line(4,3){40}}
 \put(136,39){\circle*{2}}
 \put(135,33){\scriptsize D}
 \put(84,0){\line(-4,-3){10}}
 \put(140,42){\line(-3,4){12}}
 \put(128,58){\circle*{2}}
 \put(131,55){\scriptsize B}
 \put(128,58){\line(0,-1){16}}
 \multiput(128,40)(0,-8){4}{\line(0,-1){4}}
 \put(128,8){\line(0,-1){10}}
 \put(128,43){\circle*{2}}
 \put(122,36){\scriptsize C}
 \put(90,30){\line(3,1){80}}
 \put(108,32){\scriptsize$\alpha$}
 \put(105,37){\scriptsize$\delta$}
 \end{picture}
\hfill\break
\hfill\break
\hfill\break
\caption{On the barycentric correction}
\label{BaryFig}
\end{figure*}
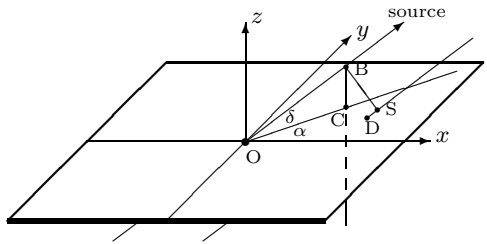

\newpage

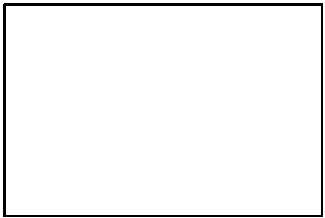
\begin{figure*}[h]
\rule{50mm}{0pt}
 \begin{picture}(270,80)
  \put(0,0){\line(1,0){120}}
  \put(120,0){\line(0,1){80}}
  \put(120,80){\line(-1,0){120}}
  \put(0,80){\line(0,-1){80}}
 \end{picture}
\hfill\break
\hfill\break
\hfill\break
\caption{On the absolute maximum structure}
\label{MaxStructure}
\end{figure*}

\newpage

\begin{figure*}[h]
\rule{50mm}{0pt}
 \begin{picture}(270,205)
 \multiput(0,0)(10,0){19}{\line(0,1){180}}
 \multiput(0,0)(0,10){19}{\line(1,0){180}}
 \put(153, 35){\circle*{2}} \put(155, 32){\scriptsize 1}
 \put( 73,120){\circle*{2}} \put( 75,122){\scriptsize 2}
 \put(118,90){\circle*{2}} \put(110,83){\scriptsize 8,3}
 \put(131,87){\circle*{2}} \put(132,88){\scriptsize 4}
 \put(131,86){\circle*{2}} \put(132,82){\scriptsize 5}
 \put(125,90){\circle*{2}} \put(125,92){\scriptsize 6}
 \put(118,96){\circle*{2}} \put(113,94){\scriptsize 7}
 \put( 80,190){$97^\circ44^\prime$}
 \put(-50, 88){$+17^\circ48^\prime$}
 \put(187, 88){\bf Epoch 1950}
 \put(-14, 18){\scriptsize $26^{\prime\prime}$}
 \put(-14, 38){\scriptsize $28^{\prime\prime}$}
 \put(-14, 58){\scriptsize $30^{\prime\prime}$}
 \put(-14, 78){\scriptsize $32^{\prime\prime}$}
 \put(-14, 98){\scriptsize $34^{\prime\prime}$}
 \put(-14,118){\scriptsize $36^{\prime\prime}$}
 \put(-14,138){\scriptsize $38^{\prime\prime}$}
 \put(-14,158){\scriptsize $40^{\prime\prime}$}
 \put(  6,182){\scriptsize $58^{\prime\prime}$}
 \put( 26,182){\scriptsize $56^{\prime\prime}$}
 \put( 46,182){\scriptsize $54^{\prime\prime}$}
 \put( 66,182){\scriptsize $52^{\prime\prime}$}
 \put( 86,182){\scriptsize $50^{\prime\prime}$}
 \put(106,182){\scriptsize $48^{\prime\prime}$}
 \put(126,182){\scriptsize $46^{\prime\prime}$}
 \put(146,182){\scriptsize $44^{\prime\prime}$}
 \put(166,182){\scriptsize $42^{\prime\prime}$}
 \end{picture}
\hfill\break
\hfill\break
\hfill\break
\caption{On the Geminga position}
\label{GemFig}
\end{figure*}
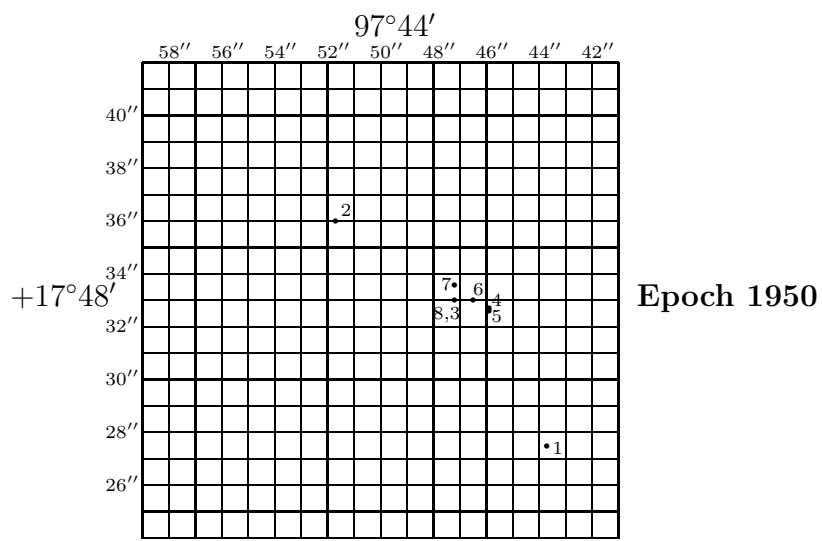

\end{document}